\documentclass[11pt]{article}

\usepackage{paspconf}
\usepackage{epsfig}
\usepackage{natbib} \bibpunct{(}{)}{;}{a}{}{;}

\newcommand{\etal}{\textit{et~al.}}

\newcommand{\hmpc}{\ensuremath{\:h^{-1}\:\mbox{Mpc}}}
\newcommand{\twod}{two-dimensional}
\newcommand{\tred}{three-dimensional}
\newcommand{\flux}{\ensuremath{\mathcal{F}}}
\newcommand{\lum} {\ensuremath{\mathcal{L}}}
\newcommand{\sect}[1]{\mbox{\S \ref{#1}}}
\newcommand{\Fig}[1] {Fig.~\ref{#1}}
\newcommand{\beeq}{\begin{equation}}
\newcommand{\eneq}{\end{equation}}
\renewcommand{\Box}{\rlap{$\sqcap$}$\sqcup$}
\newcommand{\astroref}  [1]{\texttt{(astro-ph/#1)}}
\newcommand{\definition}[1]{{\sffamily \slshape #1}}

\newcommand{\arcdeg}{^\circ}                 
\newcommand{\hour}  {\mbox{$^{\rm h}$}}      

\newcommand{\iras}{\textit{IRAS}}
\newcommand{\bootes}{Bo\"otes}
\newcommand{\potent}{\textsc{potent}}
\newcommand{\wb}{\textsc{wall builder}}
\newcommand{\vf}{\textsc{void finder}}  
\newcommand{\rmax}{\ensuremath{r_{{\rm max}}}}
\newcommand{\dmax}{\ensuremath{d_{{\rm max}}}}
\newcommand{\rvol}{\ensuremath{r_o}}
\newcommand{\di}   {\ensuremath{d_i}}
\newcommand{\dstop}{\ensuremath{d_{{\rm stop}}}}
\newcommand{\npoisson}{\ensuremath{N_{{\rm Poisson}}(d)}}
\newcommand{\nsurvey} {\ensuremath{N_{{\rm survey}}(d)}}

\begin{document}

\title{The ($>$ Half) Empty Universe}

\author{Hagai El-Ad}
\affil{Racah Institute of Physics, The Hebrew University, Jerusalem, 
       91904 Israel}

\begin{abstract}
Voids are the most prominent feature of the large-scale structure of the 
universe. Still, they have been generally ignored in quantitative analysis of 
it, essentially due to the lack of an objective tool to identify the voids 
and to quantify them. To overcome this, we have developed the \vf\ algorithm, 
a novel tool for objectively quantifying voids in the galaxy distribution. 
We apply the algorithm to two redshift surveys, the dense SSRS2 and the 
full-sky \iras\ 1.2~Jy. Both surveys show similar properties: $\sim 50\%$ of 
the volume is filled by the voids. The voids have a scale of at least 
40\hmpc, and an average under-density of $-0.9$\@. Faint galaxies do not fill 
the voids, but they do populate them more than bright ones. These results 
suggest that both optically and \iras\ selected galaxies delineate the same 
large-scale structure. Comparison with the recovered mass distribution 
further suggests that the observed voids in the galaxy distribution 
correspond well to under-dense regions in the mass distribution. This 
confirms the gravitational origin of the voids.
\end{abstract}


\section{Introduction}
\label{elad:sec:Introduction}

Perhaps one of the most intriguing findings of dense and complete nearby 
redshift surveys has been the discovery of large voids on scales of 
$ \sim 50 \hmpc $, and that such large voids appear to be a common feature of 
the galaxy distribution. Early redshift surveys like the Coma/A1367 survey 
\citep{gt78} and the Hercules/A2199 survey \citep{crt81} gave the first 
indications for the existence of voids, each revealing a void with a diameter 
of $ \sim 20 \hmpc $. Surprising as these findings may have been, it was not 
before the discovery of the \bootes\ void \citep{kir81} that the voids caught 
the attention of the astrophysical community (for a review, see 
\citealt{rood88}).

The unexpectedly large void found in the \bootes\ constellation, confirmed to 
have a diameter of $ \sim 60 \hmpc $ \citep{kir87}, brought up the question 
whether the empty regions we observe are a regular feature of the 
distribution of galaxies, or rather rare exceptions. Wide-angle yet dense 
surveys, initially \twod\ and more recently \tred, probing relatively large 
volumes of the nearby universe, established that the voids are indeed a 
common feature of the large-scale structure (LSS) of the universe. The 
publication of the first slice from the CfA redshift survey \citep{lap86} 
introduced the picture of a universe where the galaxies are located on the 
surfaces of bubble-like structures, with diameters in the range 25--50\hmpc. 
The extensions of the CfA survey \citep{gh89}, complemented in the south 
hemisphere by the SSRS and its extension, the SSRS2 \citep{dc88,dc94} have 
shown that not only large voids exist, but more importantly---that they occur 
frequently (at least judging by eye), suggesting a compact network of voids 
filling the entire volume.

The size of the structures observed in the redshift surveys is comparable to
their effective depth. With voids as large as 60\hmpc\ in diameter (or perhaps 
even larger---see \citealt{bro90}, and more recently \citealt{en97}), and 
walls extending over $ \sim 100 \hmpc $, we might still not be seeing the full 
scope of the inhomogeneities in the distribution of galaxies, as we are 
limited by the surveys' dimensions. The largest survey available today, the 
\emph{Las Campanas Redshift Survey} (LCRS) shows that structures on the scale 
of 100\hmpc\ are a common feature in the local ($ z \leq 0.2 $) universe 
\citep{lan96}. The LCRS, having an effective depth of 400\hmpc\ (if still only 
a \twod\ survey), may thus suggest that we have reached the scale where the 
universe becomes homogeneous, as was speculated earlier \cite[e.g.,][]{lap94}. 
On the other hand, results from the shallower, though \tred\ SSRS2 and CfA2 
surveys \citep{dc94} indicate that we have not yet reached a fair sample. 
This question will be resolved only through the new generation of automated 
redshift surveys, like the \emph{2-degree-Field} (2dF) survey \citep{lv96} and 
the \emph{Sloan Digital Sky Survey} (SDSS), expected to include up to $10^6$ 
galaxies \citep{love96}---compared to today's $\sim 10^4$ galaxy surveys. 
These new surveys should be completed during the first decade of the next 
century.

It has been recognized early on that inhomogeneities on such scales could 
impose strong constraints on theoretical models for the formation of LSS\@. 
However, the voids have been largely ignored and their incorporation into 
theories of LSS has been relatively recent \citep{blu92,du93,pi93}\@. The 
major obstacle here has been the difficulty of developing proper tools to 
identify and to quantify them in an objective manner.
As such, the description of a void-filled universe with a characteristic scale
of 25--50\hmpc\ relied solely on the visual impression of redshift maps. In 
order to make a more quantitative analysis we have developed the \vf\ 
algorithm \citep{ep3} for the automatic detection of voids in \tred\ surveys. 
Unlike other statistical measures, our target is to identify 
the \emph{individual voids}, in as much the same way as voids are identified 
by eye. The main features of the algorithm are: 
\begin{enumerate}
\item
It is based on the point-distribution of galaxies, without introducing 
any smoothing scale which destroys the sharpness of the observed features.
\item
It allows for the existence of some galaxies within the voids, recognizing 
that voids need not be completely empty.
\item
It attempts to avoid the artificial connection between neighboring voids
through small breaches in the walls, realizing that walls in the galaxy
distribution need not be homogeneous as small-scale clustering will
always be present.
\end{enumerate}

After a review of some of the methods for analyzing the LSS 
(\sect{elad:sec:Background}), we describe the \vf\ algorithm 
(\sect{elad:sec:Algorithm}) and the way it was tested using Voronoi 
tessellations (\sect{elad:sec:Voronoi}). We then apply the algorithm to the 
SSRS2 redshift survey (\sect{elad:sec:SSRS2}) and to the \iras\ 1.2~Jy 
(\sect{elad:sec:IRAS}). Finally (\sect{elad:sec:Discussion}), we discuss the 
results and summarize them.


\section{Background}
\label{elad:sec:Background}

The various methods for describing the void content of the LSS of the 
universe can be divided into two categories: statistical measures, and
algorithms for identifying individual voids within a sample. Additional
\emph{topological} measures, like the genus curve or percolation analysis, 
will not be discussed here.

The major statistical tool used for describing the voids is the 
\emph{Void Probability Function} (VPF). It measures the probability $P_0(V)$
that a randomly positioned sphere of volume $V$ contains no galaxies
\citep{wt79}.
For a completely uncorrelated (Poissonian) distribution, it is
$ P_0(V) = \exp(-nV) $,
where $n$ is the number-density of galaxies, so
that any departure from this quantity represents the signature for
the presence of clustering.
The major drawback of the VPF is that it is very sensitive to the
details of the galaxy distribution. For instance, adding a few galaxies in
the under-dense regions may greatly modify the VPF. A less sensitive
variant of the VPF is the \emph{Under-dense Probability Function} (UPF),
defined as the probability $ P_{\delta \rho / \rho}(V) $ that a randomly 
positioned sphere of volume $V$ has a $ \delta \rho / \rho $ under-density 
\citep{lw94}.

The first zero-crossing of the two-point correlation function $\xi(r)$ was 
used by \cite{gol95} for determining the maximum diameters of voids in case
of spherical voids. The two-point correlation function is defined as the
probability in excess of Poisson distribution of finding a galaxy in a 
volume $\delta V$ at a distance $r$ away from a randomly chosen galaxy:
\beeq  \delta P = n \, \delta V \, [ 1 + \xi(r) ]  \eneq 
where $n$ is the mean galaxy number-density. For a cellular like distribution 
the first-zero crossing is a direct measure of the characteristic size of the 
cells. Using Voronoi tessellations they show that despite the large 
uncertainty in the determination of $\xi(r)$ on large scales ($ > 20 \hmpc $), 
the zero-crossing statistic may be a useful tool in determining the scale of 
typical voids---if the galaxy distribution is void-filled, and if there is a 
characteristic scale for the void distribution. Examining the SSRS2 sample, 
they found that $ R_{\rm zero} \approx 38 \hmpc $.

Previous works have used various definitions for voids, and applied 
different algorithms to identify them.
Perhaps the first work identifying voids in a quantitative manner is
that of \cite{pel89}, who examined ensembles of contiguous cells 
with densities below a given threshold.
They use a cubic lattice, and define a local density
for each cell in the lattice. This local density is based on the analysis of
the smoothed density field. Groups of cells with densities below a specific
limit constitute the voids. The algorithm considers
two cells as contiguous if they are in contact either through their
faces, edges or vertices. This technique was applied to the original
SSRS, identifying 4 to 8 voids, depending on the density threshold. 
The major shortcomings of this algorithm are its use of a smoothed density 
field, and the lack of sense of the shape of the void it recognizes,
allowing for practically any void shape.

\cite{kf91} designed a more elaborate algorithm. They too used (empty) cubes, 
to which adjacent faces are attached. However, in order to avoid long 
finger-like extensions leading from one void into other voids, they impose a 
constraint on the adjacent faces, that each face must have an area of no less 
than two-thirds that of the surface on to which it is to be added. This scheme 
is restrictive, as it is tailored for finding only ellipsoidal-shaped voids. 
The algorithm was applied to the \emph{Southern Redshifts Catalog} (SRC) and 
to an all-sky catalog, finding a peak in the spectrum of void diameters 
between 8 and 11\hmpc. This result is inconsistent with other estimates of the 
void sizes.

A more recent work applying another void search algorithm, is that of 
\cite{lin95}. In this work single spheres, that are devoid of a certain type 
of galaxies (depending on the morphological type and luminosity), are used. 
The algorithm was applied to an area north of the super-galactic (SG) plane, 
showing that voids defined by bright elliptical galaxies have a mean diameter 
of up to 40\hmpc, in agreement with the \vf\ results. When considering 
fainter galaxies, the voids are smaller, with the faintest galaxies studied 
defining 8\hmpc\ voids, suggesting that faint galaxies delineate smaller 
voids within larger ones, which are defined by the bright galaxies.


\section{The {\sc Void Finder} Algorithm}
\label{elad:sec:Algorithm}

The \vf\ algorithm \citep{ep3} was designed with the following conceptual 
picture in mind: The main features of the LSS of the universe are voids 
surrounded by walls. 
The \emph{walls} are generally thin, \twod\ structures characterized by a high 
density of galaxies. They constitute boundaries between under-dense regions, 
generally ellipsoidal in shape---the \emph{voids}.
Although coherent over large scales, the walls---being subject to 
small-scale clustering---are not homogeneous and contain small breaches which
we wish to ignore. 
Galaxies within walls are hereafter labeled \emph{wall galaxies},
while the non-wall galaxies are named \emph{field galaxies}.
The voids are not totally empty: there are a few galaxies in them, which we
call \emph{void galaxies}.

We define a void as \definition{a continuous volume that does not contain 
any wall galaxies, and is thicker than an adjustable limit}. In other words, 
one can freely move a sphere with the minimal diameter all through the 
void. This definition does not pre-determine the shape of the void: it can 
be a sphere, an ellipsoid, or have a more complex shape, including
a non-convex one. The definition is targeted at identifying the same regions
that would be recognized as voids, when interpreting a point distribution
by eye. As the voids are defined based on the point distribution of 
galaxies, we do not need to introduce any smoothing scale. 
Our voids may contain galaxies. A stiffer requirement, such that voids should 
be completely empty, is too restrictive as a single galaxy located in the 
middle of what we would like to recognize as a void might prevent its 
identification. However, for this definition to 
be practical we must be able to identify the field galaxies before we 
can start locating the voids.

The algorithm is divided into two steps. First the \wb\
identifies the wall galaxies and the field galaxies. Then the \vf\
locates the voids in the wall galaxy distribution. 
We define a wall galaxy as \definition{a galaxy that has at least 
$n$ other wall galaxies within a sphere of radius $\ell$ around it}.
The radius $\ell$ is hereafter referred to as the 
\emph{wall separation distance}. It is derived based on the statistics of the
distance to the $n$'th nearest neighbor.
A galaxy that does not satisfy this definition is classified as a field galaxy.
This is a recursive definition which we apply successively until all the 
galaxies are classified.

\begin{figure}
\plotone{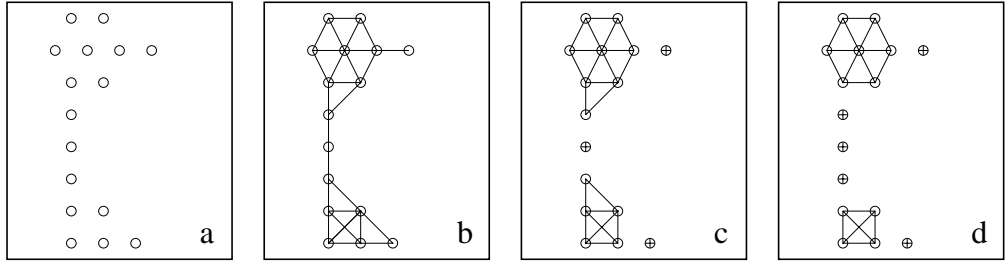}
\caption[Wall construction using the \wb]
{Wall construction using the \wb.
\emph{Panel~a}: A toy distribution of 16 galaxies ($\circ$).
\emph{Panel~b}: After the calculation of $\ell$, all galaxy pairs closer
than this separation are marked.
\emph{Panel~c}: Galaxies with less than three neighbors are flagged as field 
galaxies ($\oplus$).
\emph{Panel~d}: The final result: the string extending between the dense 
structures has been eliminated.}
\label{elad:fig:wb}
\end{figure}

\Fig{elad:fig:wb} demonstrates how the \wb\ works for $n = 3$. Notice how the 
galaxy string is filtered, while the dense structures are identified and 
maintained. As a side-bonus of this procedure, originally dedicated to 
filtering the field galaxies, we obtain a visual identification of the walls. 
This is done by drawing all the links between wall galaxies satisfying 
$ {\mathrm{dist}}(\vec x_i, \vec x_j) < \ell $. These connections are not used 
in the next step of void analysis, but provide us with another visual tool to 
examine our results (e.g., see \Fig{elad:fig:Voronoi})\@.

The \vf\ initially locates the voids containing the largest empty spheres. 
Following iterations locate the smaller voids, and---when 
appropriate---enlarge the volumes of the older voids. 
Spheres that are devoid of wall galaxies are used as building blocks for the 
voids. A single void is composed of as many superimposing spheres as required 
for covering all of its volume. The algorithm is iterative, with subsequent 
iterations searching for voids using a finer \emph{void resolution}, defined 
as the diameter \di\ of the minimal sphere used for encompassing a void 
during the $i$'th iteration. The spheres for covering a void are picked up in 
two stages: the \emph{identification stage}, followed by 
\emph{consecutive enhancements}.

During the identification stage we locate the central parts of the void. 
Usually, these spheres cover only about half of the actual volume. We focus 
(at this stage) on identifying a certain void as a separate entity, rather 
than trying to capture all of its volume. The central parts of a void are 
covered using spheres with diameters in the range $ \xi \dmax < d \leq \dmax $,
with $\dmax$ denoting the diameter of the void's largest sphere. The parameter
$\xi$ is the \emph{thinness parameter}, which controls the flexibility allowed 
at this stage.
Once a group of such intersecting spheres has been dubbed a void, it
will not be merged with any other group. 
If the void is composed of more than one sphere (as is usually the case), 
then each sphere must intersect at least another one with a circle wider than 
the minimal diameter $ \xi \dmax $. We have taken $ \xi = 0.85 $, 
which allows for enough flexibility---still without accepting 
counter-intuitive void shapes.

\begin{figure}
\plotone{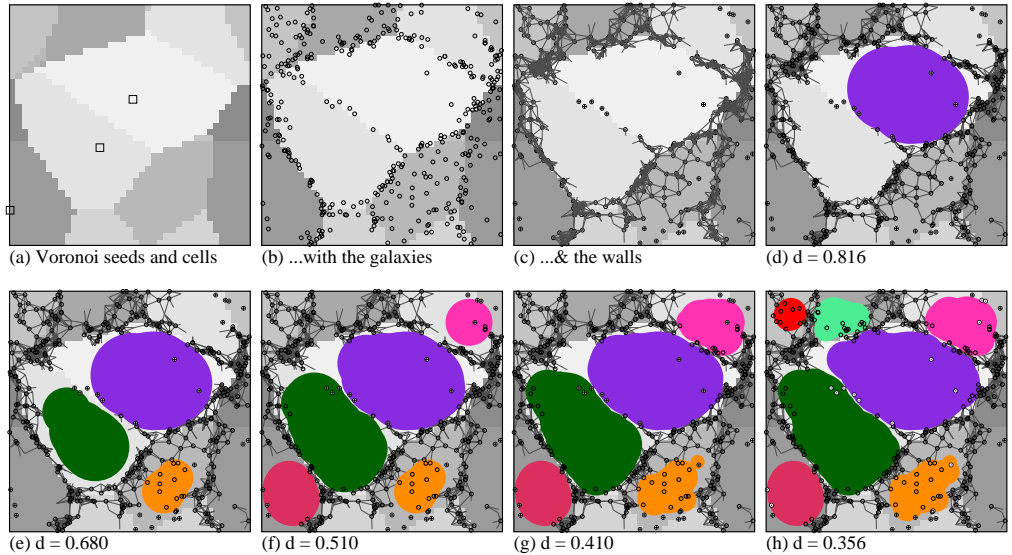}
\caption{A demonstration of the way the \vf\ covers the voids. All panels 
depict the same slice, cut through a certain Voronoi tessellation. We present 
the Voronoi seeds (\Box) \& cells (panel~a), the galaxies (panel~b) and the 
walls (panel~c). The remaining panels show the voids' image, at various void 
resolutions \di. More voids are recognized as we refine \di, and the older 
voids are enlarged.}
\label{elad:fig:vf-demo}
\end{figure}

After the central part of a void is identified, we \emph{consecutively 
enhance} its volume, in order to cover as much of the void volume as 
possible using the current void resolution. These additional
spheres need not adhere to the $\xi$ thinness limitation: 
we scan the immediate surroundings of each void, and if empty spheres are
found then they are added to the void.
We scan for enhancing spheres of a certain diameter only \emph{after}
scanning for new voids with that diameter. In this way we do not falsely
break apart individual voids, and we do not prevent the identification
of truly new voids (see \Fig{elad:fig:vf-demo}).

To assess the statistical significance of the voids we compare the voids found 
in observed data with voids found in equivalent random distributions.
The random distributions mimic the sample's geometry and 
density, and are analyzed by the algorithm in exactly the same manner.
Averaging over the random catalogs we calculate \npoisson, the expected number 
of voids in a Poisson distribution as a function of the void resolution $d$.
We compare this with the observed number, \nsurvey\@.
We define the \emph{confidence level} as:
\beeq \label{confid}
p(d) = 1 - \frac{\npoisson}{\nsurvey}
\eneq
The closer $p(d)$ is to unity, the less likely the void 
could appear in a random distribution.
We consider voids with $ p \geq 0.95 $ as statistically significant.
At a certain void resolution \dstop, \npoisson\ exceeds \nsurvey,
and we terminate the void search.

The average galaxy number-density decreases with depth in a magnitude-limited 
redshift survey. If not corrected, this selection effect will interfere with 
the algorithm in the deeper regions of the sample: field galaxies will occur 
more frequently, and the derived size of the voids will be larger. 
Consequently, systematically larger voids will be found at greater distances.
To avoid these effects, one should use a volume-limited sample, in which 
the galaxy number-density is constant and independent of the distance. A 
volume-limited sample with $ M \leq M_o $ has a depth:
\beeq 
\rvol = \sqrt{\lum_o / 4 \pi \flux_{m_b}} = 10^{-5-0.2(M_o-m_b)} \, \mbox{Mpc}
\eneq
where $m_b$ is the survey's magnitude limit and $\lum_o$ is the luminosity 
that corresponds to $M_o$.
However, current volume-limited samples are too small to study the LSS.
To overcome this, we use a semi--volume-limited sample: volume-limited up to 
some medium radius \rvol, and magnitude-limited beyond. We choose the depth 
\rvol\ by maximizing the number of bright galaxies $N(M \leq M_o)$:
\beeq  
N(M \leq M_o) = \frac{4 \pi}{3} \rvol^3  \cdot  \eta \Gamma(1-\alpha, x_o)
\eneq
where $\eta$ is the galaxy number-density. The incomplete $\Gamma$-function 
arises from the integration of the appropriate Schechter function 
\citep{sch76}, with $ x = \lum / \lum_\star $.

No corrections are needed in the volume-limited region. We determine the 
values for $\ell$ and \di\ in this region. Beyond \rvol\ we define $\phi(r)$, 
a selection-function based on the Schechter luminosity-function: 
\beeq  \phi(r) = \frac{ \Gamma(1-\alpha, x_M) }
                      { \Gamma(1-\alpha, x_{M_o}) }  \eneq
where $ x_M = 10^{-0.4(M - M_\star)} $.
The selection-function $\phi(r)$ is the observed fraction of galaxies at the 
distance $r$, relative to \rvol. 
Using $\phi(r)$ we modify both phases of the algorithm. In the 
\wb\ phase, we scale the spheres' diameters by $\phi(r)$, thus considering 
larger spheres when counting neighbors at $ r > \rvol $. 
The same correction is applied to the \vf\ phase: A void of a given size 
found in a low density environment is less significant than a void of the 
same size found in a high density environment. In order that all the voids 
found in a given iteration are equally significant, we adjust the algorithm 
so that at a given iteration relatively larger voids are accepted, if 
located at $ r > \rvol $.


\section{Voronoi Distributions}
\label{elad:sec:Voronoi}

As a test-bed for the \vf\ algorithm, we use Voronoi distributions: A
distribution of galaxies that is based on a Voronoi tessellation
\citep{vor08}.
A Voronoi tessellation is a tiling of space into convex polyhedral 
cells, generated by a distribution of seeds. 
To generate a galaxy distribution in which the galaxies are located on the 
walls of the Voronoi cells, we have used an algorithm developed by 
\cite{vdw89}. The resultant galaxy distribution has the desired 
characteristic of large empty regions (i.e., voids), which we identify by 
the \vf\ algorithm.

\begin{figure}[t!]
\plotone{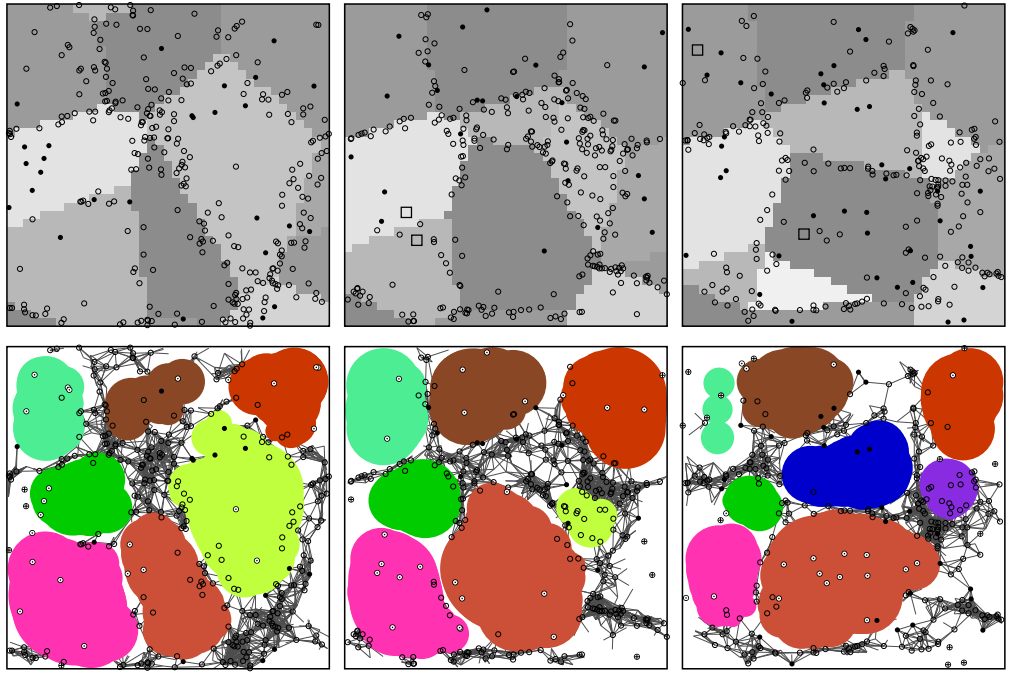}
\caption{Three consecutive slices in a Voronoi tessellation, generated from 10 
seeds with 3000 galaxies (10\% random). 
\emph{Upper row}: The Voronoi cells are depicted using gray shades, indicating 
the intersection of the cell with the central plane of the slab. Voronoi seeds 
are marked by `\Box'. Galaxies associated with cell boundaries are marked by 
`$\circ$', and random galaxies by `\textbullet'. 
\emph{Lower row}: The reconstructed voids. The voids are indicated using 
different colors, where the depicted voids correspond to the intersection of 
the central plane of each slab with the \tred\ voids. Also shown are the walls 
(dark lines marking connections between nearby galaxies). Field galaxies, 
outside the voids, are marked by `$\oplus$', and void galaxies by `$\odot$'.}
\label{elad:fig:Voronoi}
\end{figure}

A \emph{Voronoi tessellation} is constructed from a given set of 
\emph{seeds} $ \{ \vec x_i \} $.
Based on the locations of these seeds, we divide the volume into cells.
The Voronoi cell $\Pi_i$ of seed $i$ is defined by the following set of
points $\vec x$:
\beeq  \Pi_i = \{ \: \vec x \;\;\: | \;\;\: 
                  {\mathrm{dist}}(\vec x, \vec x_i) <
                  {\mathrm{dist}}(\vec x, \vec x_j) \quad \;
                  \mbox{for all $ j \neq i $} \: \}         \eneq
We assign a finite width to the walls and position the galaxies on 
the boundaries between the Voronoi cells, with a Gaussian displacement 
in the distance from the exact cell boundary. 
Additional random galaxies correspond to field galaxies.
We will call a galaxy distribution constructed in this way a \emph{Voronoi 
distribution}. The location and number of 
the Voronoi cells (the would-be voids), the spread of the wall galaxies 
and the fraction of random galaxies are all known. Therefore, we 
can use this distribution as a test bed for our algorithm.

We have constructed various Voronoi distributions and compared the original 
Voronoi tessellation to the \vf\ reconstruction (e.g., see 
\Fig{elad:fig:Voronoi}).
All Voronoi cells are reproduced except the very small cells near the
boundaries, that are cut by the box limit.
The reconstructed voids follow closely the original Voronoi cells, 
withstanding the noise introduced by the random galaxies. 
The walls highlighted using the \wb\ are located along the boundaries between 
the Voronoi cells. 
We have also created mock surveys, based on Voronoi distributions. Galaxies in 
the Voronoi distribution were assigned magnitudes according to a Schechter 
function. Then, a magnitude-limited sample was chosen. To this Voronoi-based 
mock survey we have applied our usual procedure: analyzing a 
semi--volume-limited sample and applying corrections beyond \rvol. The fit 
between the Voronoi cells and the recovered voids is still good, showing the 
adequacy of our method in analyzing actual surveys.

All together, the Voronoi tessellations that we examined show that the \vf\ 
indeed generates a faithful reproduction of the Voronoi cells. Further still, 
in cases where the reproduction merges adjacent Voronoi cells into one void, 
we see this as the adequate outcome of a missing wall. If we would have 
examined such a galaxy distribution by eye, with no prior knowledge about the 
locations of the Voronoi cells, we too would most likely consider that 
volume---originally occupied by two Voronoi cells---as one void. A level of 
$\sim 10\%$ random galaxies is tolerated, with no significant distortion in 
the void reproduction, and Voronoi-based mock surveys are also reproduced 
faithfully.


\section{The SSRS2 sample}
\label{elad:sec:SSRS2}

The SSRS2 survey \citep{dc94} consists of $\sim 3600$ galaxies with 
$ m_b \leq 15.5 $ in the region $ -40 \arcdeg < \delta < -2\fdg5 $ and 
$ b \leq -40 \arcdeg $, covering $ 1.13\, \mbox{sr} $\@. We have considered a 
semi--volume-limited sample, in this case consisting of galaxies brighter 
than $ M_o \leq -19 $, corresponding to a depth $ \rvol = 79.5 \hmpc $. The 
Schechter luminosity function was evaluated with $ M_\star = -19.5 $ and 
$ \alpha = 1.2 $, as derived for the SSRS2. Our final semi--volume-limited 
sample consists of 1898 galaxies, extending out to $ \rmax = 130 \hmpc $ 
where the selection-function $\phi$ has dropped to 17\%.
It should be emphasized that the SSRS2 analysis is performed in 
\emph{redshift-space}. However, because of the paucity of large clusters and 
the small amplitude of peculiar motions in the volume surveyed by the SSRS2, 
redshift distortions are small \citep{dc97} and the properties derived here 
should reflect those of voids in real-space.

The \wb\ analysis of the SSRS2 has classified 91.5\% of the galaxies 
as wall galaxies, and 8.5\% as field galaxies.
The wall separation distance was $ \ell = 7.4 \hmpc $. In the volume-limited 
region we have $ n^{-1/3} = 6.4 \hmpc $, so $ \ell / n^{-1/3} = 1.16 $.
The wall galaxies are grouped in ten structures: one structure contains most 
(96\%) of the wall galaxies. The rest 
of the wall galaxies are found in nine groups, each having 4 to 21 galaxies.

We have identified eleven significant ($ p \geq 0.95 $) voids within the 
volume probed by the SSRS2\@. These voids were detected while the void 
resolution was $ d \geq 19.2 \hmpc $. In the following calculations we 
take into account only these voids, unless otherwise specified. Seven 
additional voids were identified before the void search was terminated at the 
resolution $ \dstop = 15.1 \hmpc $, for which $p$ vanishes. 
Initial results for the SSRS2 were reported in \cite{epd1}.
The locations and characteristics of all eighteen voids are given in 
Table~1 of \cite{ep3}.

\begin{figure}
\plotone{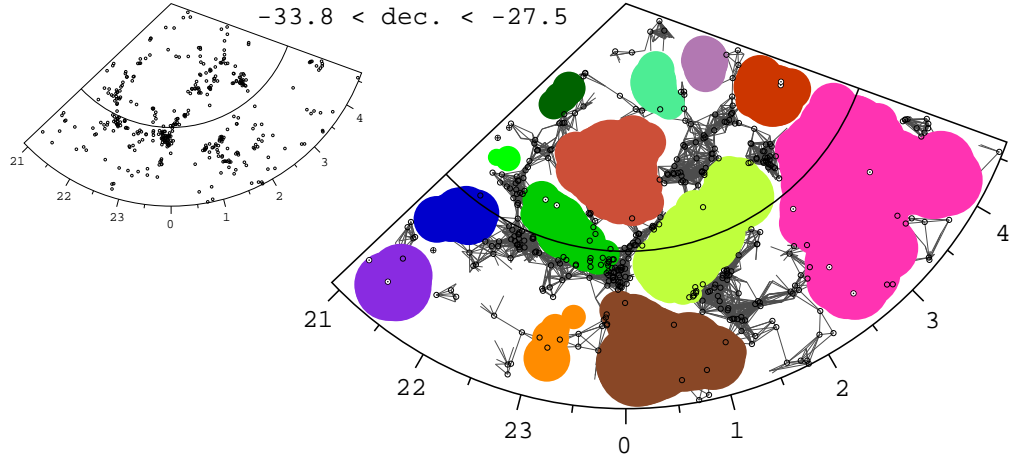}
\caption{A slice through the SSRS2 sample (\emph{left}) and the resultant \vf\ 
reconstruction (\emph{right}).
The slice contains many voids, walls and filaments.
The outstanding structure from ($ \alpha = 4\hour $, $ r = 45 \hmpc $) to 
($ \alpha = 0\hour $, $ r = 90 \hmpc $) is the Southern Wall (SW). 
The Pavo-Indus-Telescopium (PIT) supercluster runs along the line of sight at 
$ \alpha = 21\fh5 $.}
\label{elad:fig:SSRS2}
\end{figure}

The average size of the voids in the SSRS2 as estimated from the 
equivalent diameters is $ \bar{d} = 40 \pm 12 \hmpc $.
The average under-density within the voids was 
found to be $ \delta \rho / \rho \approx -0.9 $, a quite remarkable result 
showing how empty voids are of bright galaxies.
The eleven significant voids comprise 54\% 
of the survey's volume. An additional 5\% is covered by the seven
additional voids, totaling in $ \sim 60\% $ of the volume being occupied
by these voids.
We estimate that the walls occupy less than 25\% of the volume.
A single $ 6\fdg25 $-wide constant-declination slice through the SSRS2 is 
presented in \Fig{elad:fig:SSRS2}.

The largest void found in the SSRS2 survey has an equivalent 
diameter $ d = 60.8 \hmpc $, making it comparable in volume to the large void 
found in the \bootes\ \citep{kir81}. This void is an ellipsoid, whose major 
axis is perpendicular to the line of sight, located at: 
$ 80 \hmpc < r < 130 \hmpc $;
$ -25\arcdeg < \delta < -2\fdg5 $; 
$ 21\hour < \alpha < 23\fh75 $. 
This void might actually be larger, since it is bounded (in three directions) 
by the limits of the SSRS2. A second large void (with $ d = 56.2 
\hmpc $) is also comparable to the \bootes\ void.

When preparing the semi--volume-limited sample, we cast off all faint 
$ M > M_o $ galaxies in the region $ r < \rvol $. These galaxies comprise the 
bulk of the surveyed galaxy population that we are forced to ignore (the rest 
are $ r > \rmax $ galaxies, where the sample is too sparse). Although we 
cannot use these galaxies during the analysis phases, we can still try and 
benefit from them \emph{a~posteriori}: after the voids are located, we 
examine the locations of these galaxies.
Almost 61\% of the $ r < \rvol $ region is covered by voids---but only 19\% 
of the 1264 faint galaxies are found within them. Even though the \vf\ 
algorithm uses only the brighter galaxies in this region, we find that the 
faint galaxies do not fill the voids, providing an excellent verification of 
the algorithm. Still, the percentage of faint galaxies within the voids is 
significantly larger than that of the bright galaxies: only $ \sim 5\% $ of 
the bright $ M_o \leq -19 $ galaxies are contained in the voids.


\section{The \iras\ sample}
\label{elad:sec:IRAS}

The \iras\ survey contains 5321 galaxies complete to a flux limit of 1.2~Jy 
\citep{fi95}. We applied corrections for the computed peculiar velocities, to 
obtain the real-space distribution of the galaxies. In the void analysis, we 
have limited ourselves to galaxies extending out to $ \rmax = 80 \hmpc $, and 
created a semi--volume-limited sample with a depth $ \rvol = 50 \hmpc $. The 
selection function drops to 22\% at $\rmax$. The final sample consists of 1876 
galaxies, and 1531 faint galaxies were eliminated in order to create the 
volume-limited region.

The sky coverage of the \iras\ is almost complete (87.6\%),
with the galactic plane region $|b| < 5 \arcdeg $ constituting most
of the excluded zones. 
Various schemes \cite[e.g.,][]{ya91} have been used to 
extrapolate the density field to the galactic plane, but these are not 
directly applicable to our analysis. Thus, when looking for voids
we avoid the ZOA, treating it as a rigid boundary practically cutting the 
\iras\ to two halves. 
Since the ZOA cuts across voids this scheme divides some voids to two
and eliminates others. However, it is the most conservative method, and 
therefore the results for the volumes of the voids should be considered as 
lower limits.
We estimate the effect of this method by examining the opposite approach 
in which the ZOA is treated as if it is a part of the survey, applying 
no corrections. 
The ZOA is nowhere wider than the minimal void resolution used, so it does 
not create new voids by itself. 
Therefore the effect of including the ZOA is to overestimate the
size of voids near it, because it allows the merging of a couple of
voids and the expansion of other voids into the region. Still the overall 
effect on the void statistics is limited.

The \wb\ analysis of the \iras\ galaxy distribution located 95\%
of the galaxies within walls.
We find that the walls occupy at most $ \sim 25\% $ of the examined 
volume. This corresponds to an average wall over-density of at least 
\mbox{$ \delta \rho / \rho \approx 4 $}. 
Since the \iras\ sample is relatively sparse, we have considered all the 
galaxies while identifying the voids. Hence, the \iras\ voids presented below 
are \emph{completely empty}. 

Applying our most conservative approach to analyze the \iras---i.e., 
including the field galaxies and avoiding the ZOA---we have identified 
24 voids of which twelve are statistically significant at a $0.95$ 
confidence level \citep{epd2}. 
In general, some of the voids we find are smaller than their actual size---%
because of the way we treat the ZOA, or because the field 
galaxies were not removed from the analysis. Both effects imply that
our estimates of the size of voids are likely to be lower limits. 

\begin{figure}
\noindent
\begin{minipage}{.469\linewidth}
\raggedleft
{\Huge a}
\epsfig{bbllx=108, bblly=540, bburx=234, bbury=666, clip=,
        file=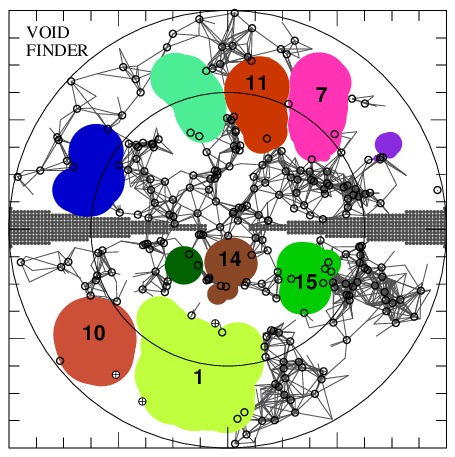, height=154pt, width=154pt}
\end{minipage}\hspace{1pt}%
\noindent%
\begin{minipage}{.469\linewidth}
\raggedright
\epsfig{bbllx=80pt, bblly=205pt, bburx=564pt, bbury=690pt, clip=,
        file=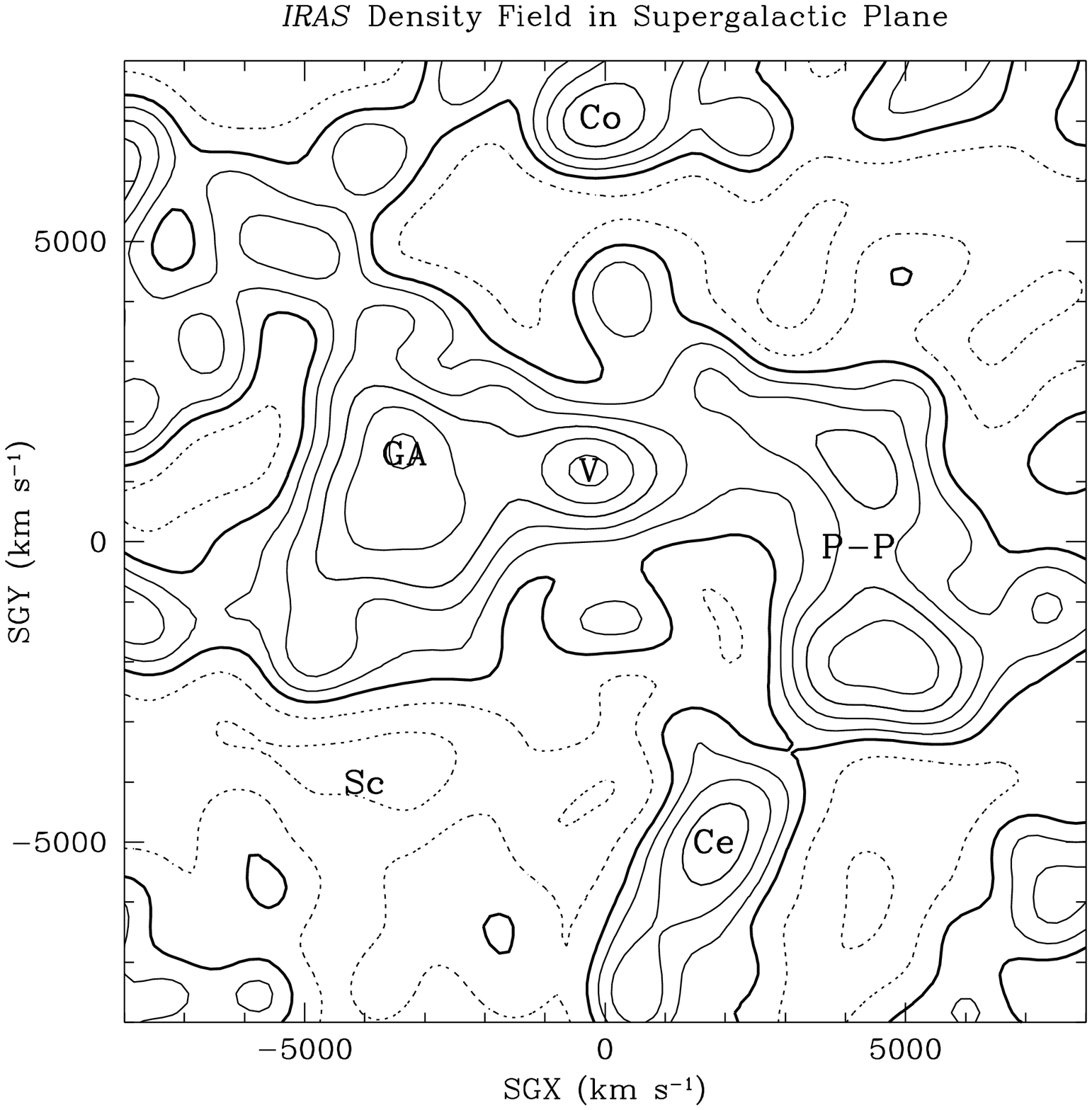, height=154pt, width=154pt}
{\Huge b}
\end{minipage}
\noindent
\begin{minipage}{.469\linewidth}
\raggedleft
{\Huge c}
\psfig{bbllx=98pt, bblly=223pt, bburx=565pt, bbury=690pt, clip=,
        file=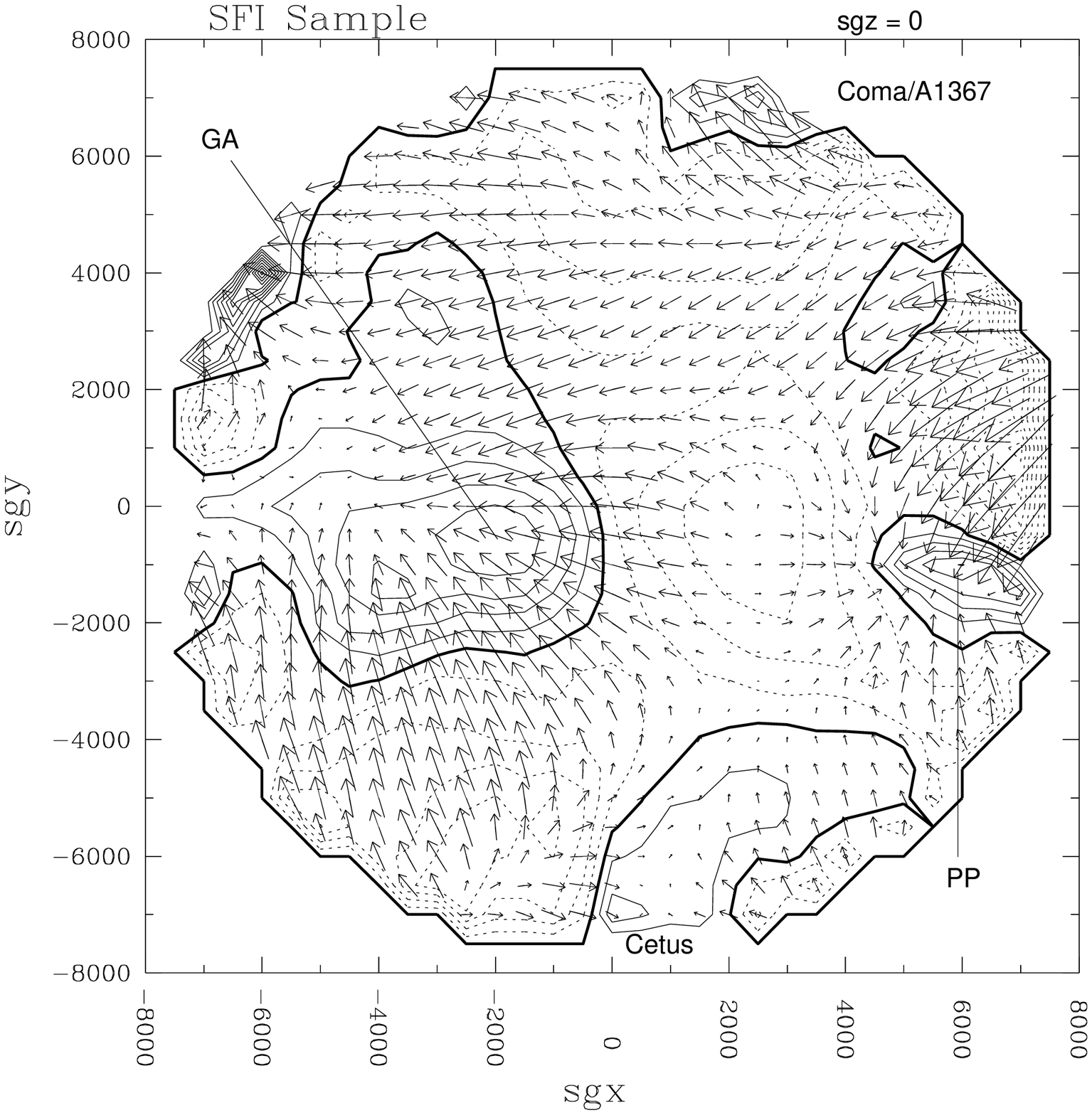, height=154pt, width=154pt}
\end{minipage}\hspace{1pt}%
\noindent%
\begin{minipage}{.469\linewidth}
\raggedright
\psfig{bbllx=84pt, bblly=174pt, bburx=529pt, bbury=620pt, clip=,
        file=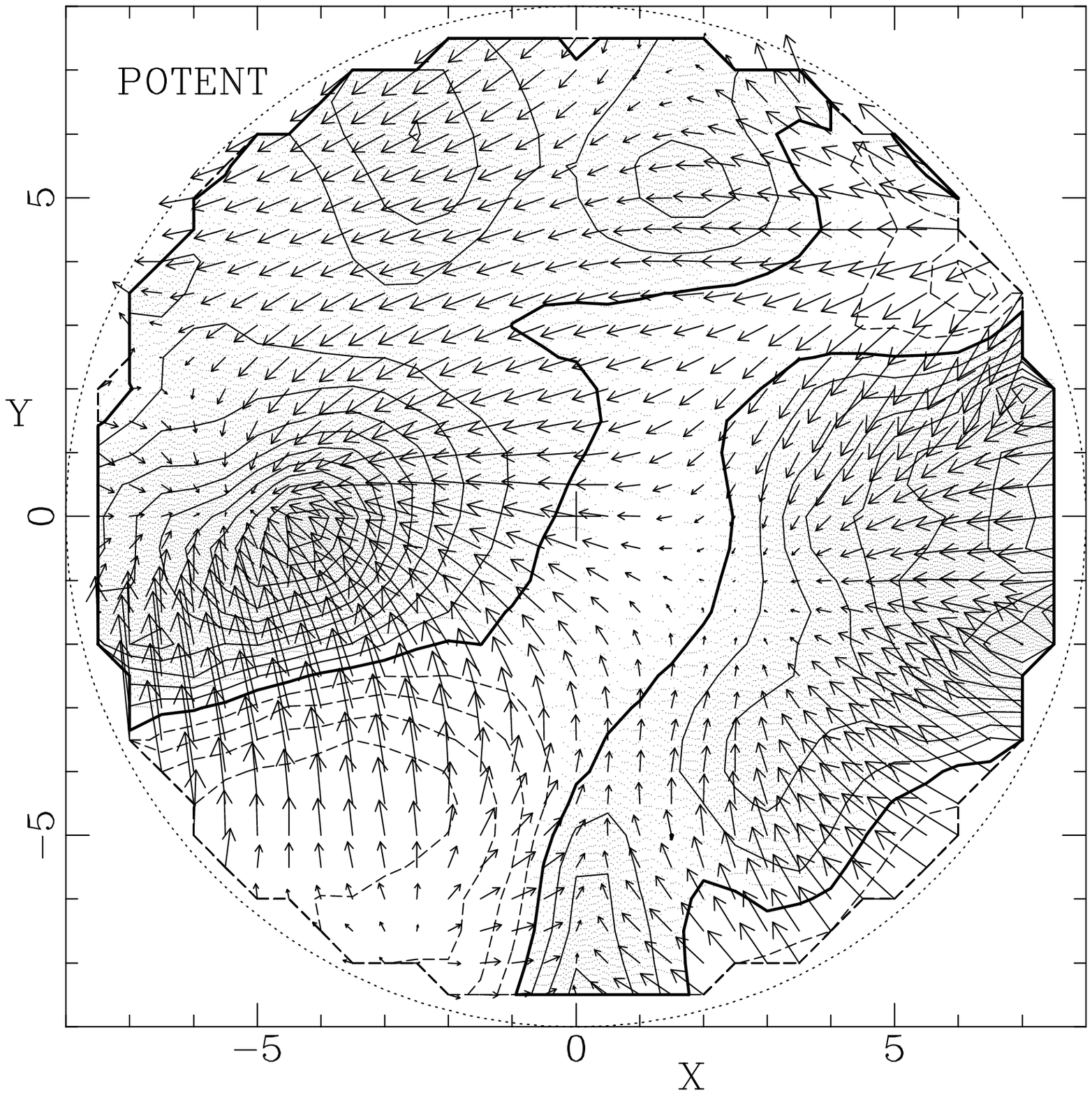, height=154pt, width=154pt}
{\Huge d}
\end{minipage}
\caption{The supergalactic plane extending out to $ 80 \hmpc $, as depicted 
by various techniques.
\emph{Panel a} \protect\citep{epd2}: The real-space locations of the voids and 
the walls in the \iras\ 1.2~Jy sample, using the \vf\ algorithm. The excluded 
ZOA is indicated along $ Y = 0 $. The depicted galaxies extend $ 5 \hmpc $ 
above and below the plane. Wall galaxies are marked as by `$\circ$', field 
galaxies by `$\oplus$'. \emph{All} the galaxies are located outside the 
voids---galaxies that seem to be in a void appear so due to the 
two-dimensional projection. The inner circle at $ \rvol = 50 \hmpc $ marks the 
volume-limited region of our sample.
\emph{Panel b} \protect\citep{sw95}: The real-space smoothed density field of 
\iras\ galaxies, using $ 5 \hmpc $ Gaussian smoothing, extrapolating into the 
ZOA\@. The density field is obtained by a self-consistent correction for 
peculiar velocities with $\beta = 1$. Reproduced by permission of Michael 
Strauss.
\emph{Panel c} \protect\citep{dc96}: The reconstructed velocity and density 
fields obtained from the SFI sample, using $ 9 \hmpc $ Gaussian smoothing. The 
arrows give the $X$--$Y$ components of the three-dimensional velocity field. 
The contours are of $\delta$, spaced at $0.2$ intervals. The heavy solid line 
indicates $\delta = 0$. Reproduced by permission of Luiz da Costa.
\emph{Panel d} \protect\citep{de94,de97}: The smoothed velocity field and the 
resultant density field as recovered by \potent\ from the Mark~III data, using 
$ 12 \hmpc $ Gaussian smoothing. Reproduced by permission of Avishai Dekel.}
\label{elad:fig:IRAS}
\end{figure}

In the SG plane (\Fig{elad:fig:IRAS}, panel {\it a\/}) one recognizes void~10 
as the Sculptor Void \citep{dc88}, located below the P-I-T part ($ Y < 0 $) of 
the Great Attractor (GA), seen here to be composed of several sub-structures.
Adjacent to it we find void~1, stretching parallel to the Cetus wall. 
These two voids are separated only by a few field galaxies. If we filter 
them out, the two would merge to form one huge void, equivalent in volume 
to a $ d = 62 \hmpc $ sphere occupying most of that part of the skies.
Voids 1 \& 10 are limited by the \rmax\ boundary of our sample, so they 
could prove to be larger still.

The area above the Perseus-Pisces (PP) supercluster (up to the Great Wall 
near Coma, at $ Y = 70 \hmpc $) is occupied by two voids: 7 \&~11. If the 
field galaxies are filtered first, these two voids merge.
Also note in this area the minor void ($ p = 0.21 $) located below the Coma
supercluster, at $( X = -7, Y = 54 )$: this void corresponds to the largest 
void found in the CfA survey \citep{lap86}.
The closest void we found (void~14), can be seen in the center of
this panel, just below the local supercluster.
Another clear, and rather nearby, void in the SG plane is void~15, 
in front of PP\@. A minor void can be viewed beyond
the $ Y > 0 $ section of the GA, at $( X = -51, Y = 19 )$. 

\begin{figure}
\plottwo{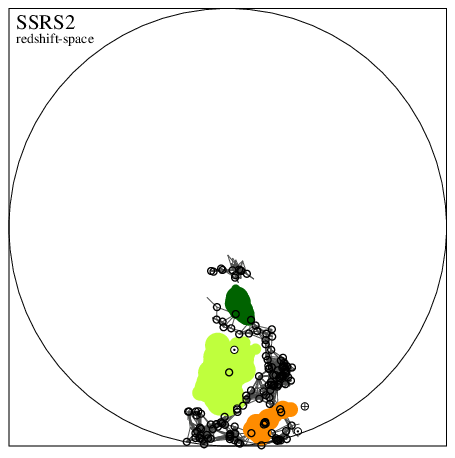}{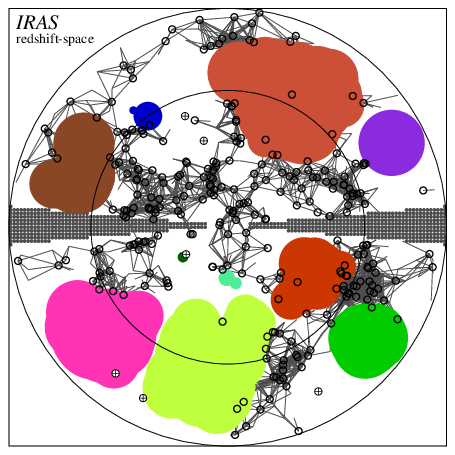}
\caption[\iras\ and SSRS2 voids in the SG plane]
{Redshift-space voids in the SG plane: SSRS2 (\emph{left}) and \iras\ 
(\emph{right}). The denser sampling of the SSRS2 is evident. Similar voids 
are found in the overlapping regions of the surveys. For the \iras, compare 
also the real- and redshift-space distribution: the \iras\ voids in 
redshift-space are larger, and the dense structures appear much more 
collapsed, than in real-space (\Fig{elad:fig:IRAS}, panel {\it a\/}).}
\label{elad:fig:IRAS-SSRS2}
\end{figure}

The average size of the twelve significant \iras\ voids as estimated from the 
equivalent diameters is \mbox{$ \bar{d} = 40 \pm 6 \hmpc $}.
The increase in average void diameter in the \iras\ compared to the 
SSRS2 ($\sim 5\%$) is due to the relatively narrow angular limits 
of the latter survey.
The twelve most significant \iras\ voids occupy 22\% of the
examined volume; considering all 24 voids, the volume is 32\%\@.
If we consider only the volume-limited
region of our sample, where there are no distortions caused by the
survey's \rmax\ boundary (only the ZOA), the void
volume reaches 46\%\@. We have also examined the void
distribution in redshift-space. As expected, voids in redshift-space
are typically bigger than their real-space counterparts. The
total void volume in redshift-space is $\sim 20\%$ larger
than that in real-space, and the average diameter of the significant 
voids in redshift-space is $ 44 \hmpc $ (compare \Fig{elad:fig:IRAS}, panel 
{\it a}, with \Fig{elad:fig:IRAS-SSRS2}, right panel).

After the voids were located we examined the locations of the previously 
eliminated faint galaxies. Only 13\% of these are located within the voids, in 
agreement with the identification of the voids based on the brighter galaxies. 
However, as found in the SSRS2, there is a notable increase in the number of 
faint galaxies in the voids, compared to the number of brighter galaxies.

What is the effect of the limitations we have imposed in our void
analysis? As stated above, the treatment of the ZOA as a rigid
boundary and the consideration of only empty voids, cause us to
interpret the results derived in this way as a lower limit. An upper
limit is derived by taking the opposite approach, this time including
the ZOA and filtering the field galaxies. Each factor alone
corresponds to an increase in the average void diameter of 5--15\%.
Together the effect is $\sim 20\%$, yielding an upper
limit for this sample of $ \bar{d} = 48 \hmpc $.
A similar increase occurs in the
total void volume. When filtering the field galaxies the voids are not
empty, now having an average under-density of
\mbox{$ \delta \rho / \rho \approx -0.9 $}, as found for the SSRS2\@.


\section{Discussion}
\label{elad:sec:Discussion}

We have used the \vf\ algorithm to analyze two redshift surveys: The SSRS2 
and the \iras\ 1.2~Jy. These surveys represent two 
extreme cases in the trade-off between density and sky coverage. The SSRS2 is 
densely sampled ($ m_b \leq 15.5 $), but it has narrow angular limits, 
especially in the declination range. As a result, voids are often limited by 
the survey's boundary, diminishing their scale. The \iras\ is an almost 
full-sky survey ($87.6\%$ coverage), but it is rather sparse. As a result one 
cannot use a small void resolution, for lack of statistical significance. In 
addition to the above differences, the SSRS2 galaxies are optically selected, 
as opposed to the \iras\ galaxies.

Withstanding these differences, the results obtained with these surveys are 
similar. First, the surveys agree regarding individual voids in the regions 
where the surveys overlap. \Fig{elad:fig:IRAS-SSRS2} depicts the 
redshift-space voids in the SG plane, for the \iras\ and for the corresponding 
part of the SSRS2\@. In the region where the SSRS2 sample overlaps the \iras\ 
sample, we find three of the eleven significant voids identified in the SSRS2. 
The corresponding \iras\ voids are $ \sim 33\% $ larger than the SSRS2 ones, 
since they are not bounded by narrow angular limits as the SSRS2 voids.

Additionally, the results agree viz a viz the voids' statistical 
characteristics:
\begin{enumerate}
\item Large voids occupy $ \sim 50\% $ of the volume.
\item Walls occupy less than $ \sim 25\% $ of the volume.
\item A void scale of at least 40\hmpc, with an average under-density of 
      $-0.9$.
\item Faint galaxies do not fill the voids, but they do populate them 
      more than bright ones. 
\end{enumerate}
The void scale derived in both surveys is a lower limit: for the SSRS2, 
because of the narrow boundaries limiting the voids; and for the IRAS, due to 
the larger \dstop, and because of the conservative analysis applied regarding 
the ZOA and the field galaxies.

The fact that both the \iras\ and the SSRS2 are consistent regarding the void 
statistics as well as the individual voids is not trivial, since the \iras\ 
galaxies represent a special galaxy class, possibly biased relative to the 
optical galaxies \citep{la88}. The agreement between the surveys suggests 
that a similar void scale exists for both optically and \iras\ selected 
galaxies. This suggests that the voids are also devoid of dark matter, 
indicating that they have formed gravitationally \citep{pi93}.

The \iras\ data also provides a suitable benchmark as it has been used to 
derive the smooth density field, and it probes a volume comparable to that 
used to determine the density field of the underlying mass distribution from 
the \potent\ reconstruction method \citep{de90}, based on the measured galaxy 
peculiar velocity field (see \Fig{elad:fig:IRAS}). The voids and walls 
identified by the \vf\ indeed correspond to the under- and over-dense 
regions in the \iras\ density field \citep{sw95} respectively. Comparison with 
the SFI sample \citep{dc96} also demonstrates that the voids delineated by 
galaxies correspond remarkably well with the under-dense regions in the 
reconstructed mass density field derived from peculiar velocities 
(but see also the Mark~III map---\citealt{de94,de97}). This confirms 
the idea suggested earlier, that the observed voids in redshift surveys 
represent true voids in the mass distribution, forcing a gravitational theory 
for their formation. 

Most of the over-dense regions, walls and filaments, are narrower than
10\hmpc. The smoothing scale used for creating the density
fields spreads the originally thin structures over wider regions,
extending into the under-dense volumes. This has the effect of giving
a false impression of a rather blurred galaxy distribution, where
prominent over-dense structures are separated by small under-dense
regions. The true picture is very different: there is a sharp
contrast between the thin over-dense structures which occupy only the
lesser part of the volume, and the large voids. The notion of a void
filled universe cannot be avoided in this picture. 

We have developed and tested a new tool for quantifying the 
large-scale structure of the universe. Unlike most of the work in this
field, we focus on the under-dense regions, and for the first time are
capable of individually identifying and statistically quantifying the voids.
The \vf\ analysis clearly shows the prominence of the voids in the LSS, 
not hindered by smoothing of the over-dense regions, and it reveals the
image of a void-filled universe, where large voids are a common feature.

The consistency in the void image between \iras\ and optically selected 
galaxies suggests that galaxies of different types delineate equally well 
the observed voids. Therefore galaxy biasing is an unlikely mechanism for 
explaining the observed voids in redshift surveys. Comparison with the 
recovered mass distribution further suggests that the observed voids in 
the galaxy distribution correspond well to under-dense regions in the mass 
distribution. This confirms the gravitational origin of the voids.

\acknowledgments
I am indebted to Tsvi Piran for his guidance and encouragement throughout this 
research.
I would like to warmly thank Luiz da~Costa for providing the SSRS2 data and 
for numerous helpful discussions.
I am grateful to Avishai Dekel and Michael Strauss, for providing 
two of the figures, and to Rien Van de Weygaert for providing his Voronoi 
tessellation code.
I am happy to acknowledge the support of the Israeli MFA and of the Casa96 
organizing committee.


\end{document}